\documentclass{sf2a-conf2025}
\usepackage{graphicx}
\usepackage{hyperref}
\usepackage[]{natbib}  
\usepackage{epstopdf}
\usepackage{xcolor} 

\def\BibTeX{{\rm B\kern-.05em{\sc i\kern-.025em b}\kern-.08em
    T\kern-.1667em\lower.7ex\hbox{E}\kern-.125emX}}
\bibpunct{(}{)}{;}{a}{}{,}  

\begin{document}

\TitreGlobal{SF2A 2025}


\title{Exploring hydrodynamical stellar tachoclines along stellar evolution}

\runningtitle{Hydrodynamical stellar tachoclines}

\author{C. Moisset}\address{Université Paris-Saclay, Université Paris Cité, CEA, CNRS, 91191 Gif-sur-Yvette, France}
\author{S. Mathis\,$^1$}
\author{L. Amard}\address{Department of Astronomy, University of Geneva, Chemin des Maillettes 51, CH-1290 Versoix, Switzerland}




\setcounter{page}{237}


\maketitle


\begin{abstract}
Stellar tachoclines are thin regions located between the radiative core and the convective envelope of solar-type stars. They are defined as layers where the rotation of the radiative interior transitions to the differential rotation of the convective envelope, generating strong shear and turbulence. As such, understanding the dynamics of the transport and mixing inside stellar tachoclines would shed light on how the dynamical processes of the convection zone might affect the secular transport of the radiative zone.
In particular, we investigate how the change of the latitudinal differential rotation in the convection zone with stellar evolution would affect the dynamics of the tachocline. Indeed, as solar-type stars are braked on the Main Sequence, the differential rotation in the convection zone is expected to evolve from a cylindrical rapidly-rotating regime (columns of varying velocities, aligned with the rotation axis) to a conical solar-like regime (with an equatorial acceleration as in the case of the Sun) and finally to a conical anti-solar-like regime (with a polar acceleration). 
However, stellar evolutionary codes currently only consider at best the solar conical regime to study the dynamics of stellar tachoclines throughout the evolution of stars. We discuss different possibilities to model hydrodynamical tachoclines and we show that Mathis \& Zahn 2004’s formalism is able to treat coherently hydrodynamical stellar tachoclines when taken in the thin layer approximation. We use it to model the differential rotation, meridional circulation, and mixing coefficients inside the tachocline in order to examine the effect of the different rotation regimes on the transport.

\end{abstract}



\section{Introduction}
The first model of tachocline was proposed by \cite{SpiegelZahn1992} who sought to explain the observations of the internal rotation of the Sun \citep[e.g.][]{Garciaetal2007}. Helioseismology indeed revealed that the solar rotation changes abruptly (i.e. over less than $5\%$ R$_{\odot}$) from the differential rotation of the convective envelope to the quasi solid-body rotation of the radiative interior. As such, the solar tachocline is a strongly sheared layer, both in the radial and latitudinal directions. As a consequence, it influences the transport of angular momentum and mixing of chemical elements in stars and has consequences over evolutionary timescales \citep[e.g.][]{Brunetal1999, Dumontetal2021}.

One of the main challenges lies in understanding the thinness of the tachocline. Indeed, the heat diffusion would normally act to spread the latitudinal shears of the convective region into the radiative interior, unless another transport mechanism acts to confine the tachocline to its current thinness. Numerous possibilities have been considered, most of them involving either hydrodynamical turbulence of the radiative region \citep[e.g.][]{SpiegelZahn1992, Garaud2025} or magnetic fields, either in the radiative \citep[e.g.][]{GoughMcIntyre1998, Strugareketal2011, AcevedoArreguinetal2013} or in the convective region \citep[e.g.][]{Barnabeetal2017, Matilskyetal2024}. In particular, the solar tachocline by \cite{SpiegelZahn1992} is an hydrodynamical model which consider that a strong anisotropic horizontal turbulence inside the tachocline would ensures its confinement. 

\begin{figure}[ht!]
\centering
\begin{minipage}[b]{0.48\linewidth}
 \centering
 \includegraphics[width=\textwidth,clip]{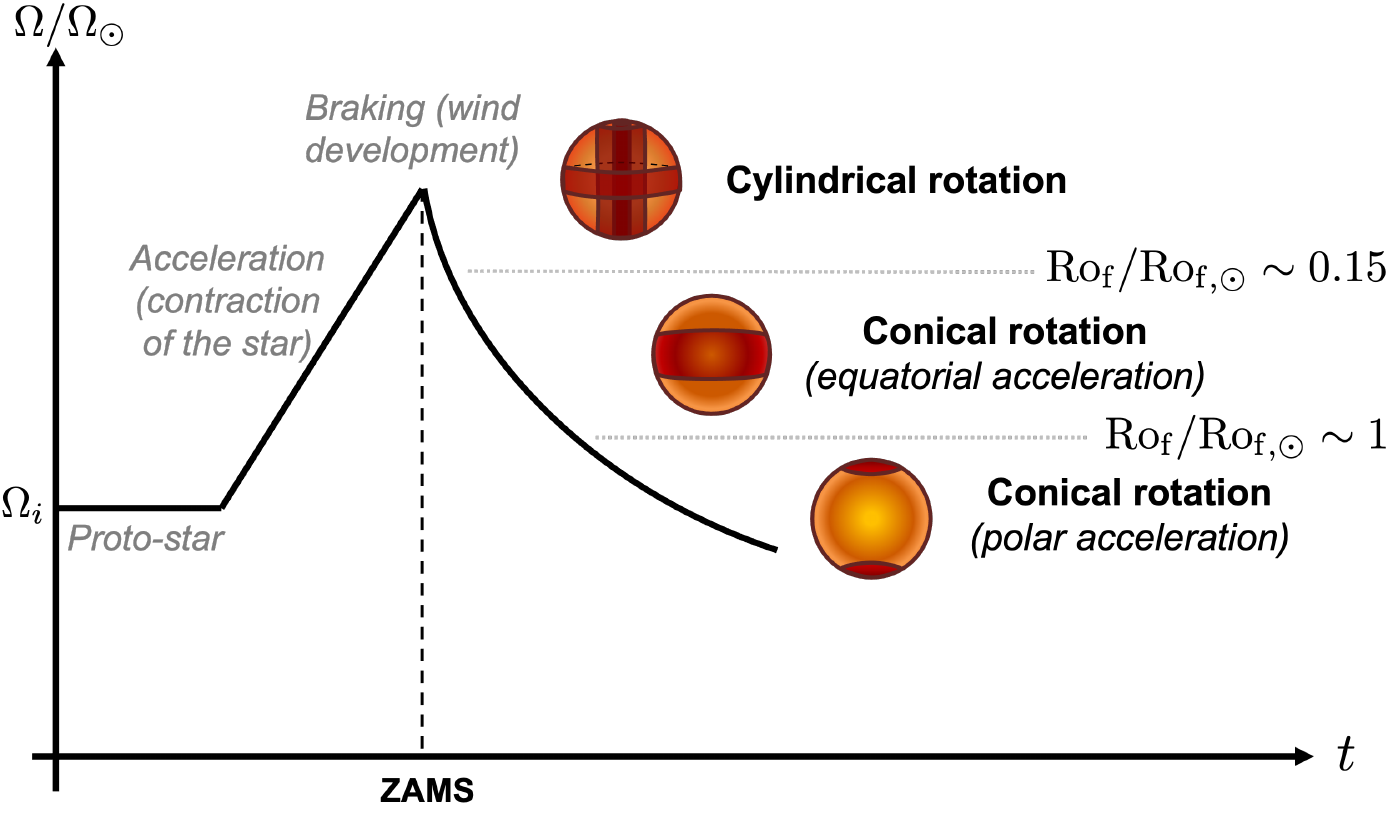}
\end{minipage}
 \quad
\begin{minipage}[b]{0.35\linewidth}
\centering
  \caption{Illustration of the temporal evolution of the mean surface rotation $\Omega$ of solar-type stars, normalised to that of the Sun $\Omega_{\odot}$. $\Omega_{i}$ is the rotation of the initial proto-star. After the ``Zero Age Main Sequence'' (ZAMS), the differential rotation in the convective envelope evolves from cylindrical to conical solar-like and conical anti solar-like, at values of the Rossby fluid number $\textrm{Ro}_{\textrm{f}}$ normalized by the solar value $\textrm{Ro}_{\textrm{f},{\odot}}$ around $0.15$ and $1$ respectively \citep[][]{Brunetal2022,Norazetal2024}.}
  \label{fig_context}
  \vspace{0.9 cm}
\end{minipage}
\end{figure}

Because stellar tachoclines are located at the borders of the radiative and convective zones, they are essential to understand the interactions between both regions. In particular, 3D MHD simulations of the convective envelope of low-mass stars have shown that the differential rotation changes during the main sequence (MS) along with the fluid Rossby number $\textrm{Ro}_{\textrm{f}}$ which measures the impact of the rotation over the convective motions \citep[][]{Brunetal2017, Brunetal2022, Norazetal2024}. As illustrated in figure \ref{fig_context}, during the early MS when the star is rotating faster, the differential rotation has a cylindrical profile (cylinders of constant velocity aligned with the axis of rotation because of the Taylor-Proudman constrain). Later on the MS, the rotation becomes conical solar-like (with a fast equator and slower poles), before transitioning to conical anti-solar-like (with slow equator and fast poles). Observations of such latitudinally sheared profiles have been provided for the cylindrical and conical solar-like rotations \citep[e.g.][]{Benomaretal2018, Bazotetal2019} and \cite{Norazetal2022} has provided some targets for the forthcoming PLATO mission for the conical anti-solar-like case. However, current stellar evolutionary codes, at best, include tachocline models forced by a conical solar-like differential rotation from the convective envelope.

Our work investigates the dynamics of stellar tachoclines over evolutionary timescales, using the hydrodynamical approach of \cite{SpiegelZahn1992}, treated here with the formalism for stellar evolution by \cite{MathisZahn2004} where we take into account the change of differential rotation in the upper convective envelope along the evolution of stars.

\section{Differential rotation in the tachocline}

To provide a self-consistent model of hydrodynamical tachocline for stellar evolution, we first demonstrated that the formalism to describe the transport of angular momentum and the mixing of chemicals in stably stratified radiative zones established by \cite{MathisZahn2004} effectively allows us to recover the hydrodynamical model by \cite{SpiegelZahn1992}, when used within the framework of a thin boundary layer. The model was then extended for different rotation profiles imposed by the overlying convective layer (either a conical or a cylindrical profile). Furthermore, we took into account the evolution of the star and used a STAREVOL model of a 1 M$_{\odot}$ star with a high initial rotation rate (which reaches up to 100 $\Omega_{\odot}$ at $t=3\times10^{7}$ years) \citep[e.g.][]{Palaciosetal2006, Amardetal2019}. 

\begin{figure}[ht!]
	\centering
	\includegraphics[width=0.7\textwidth,clip]{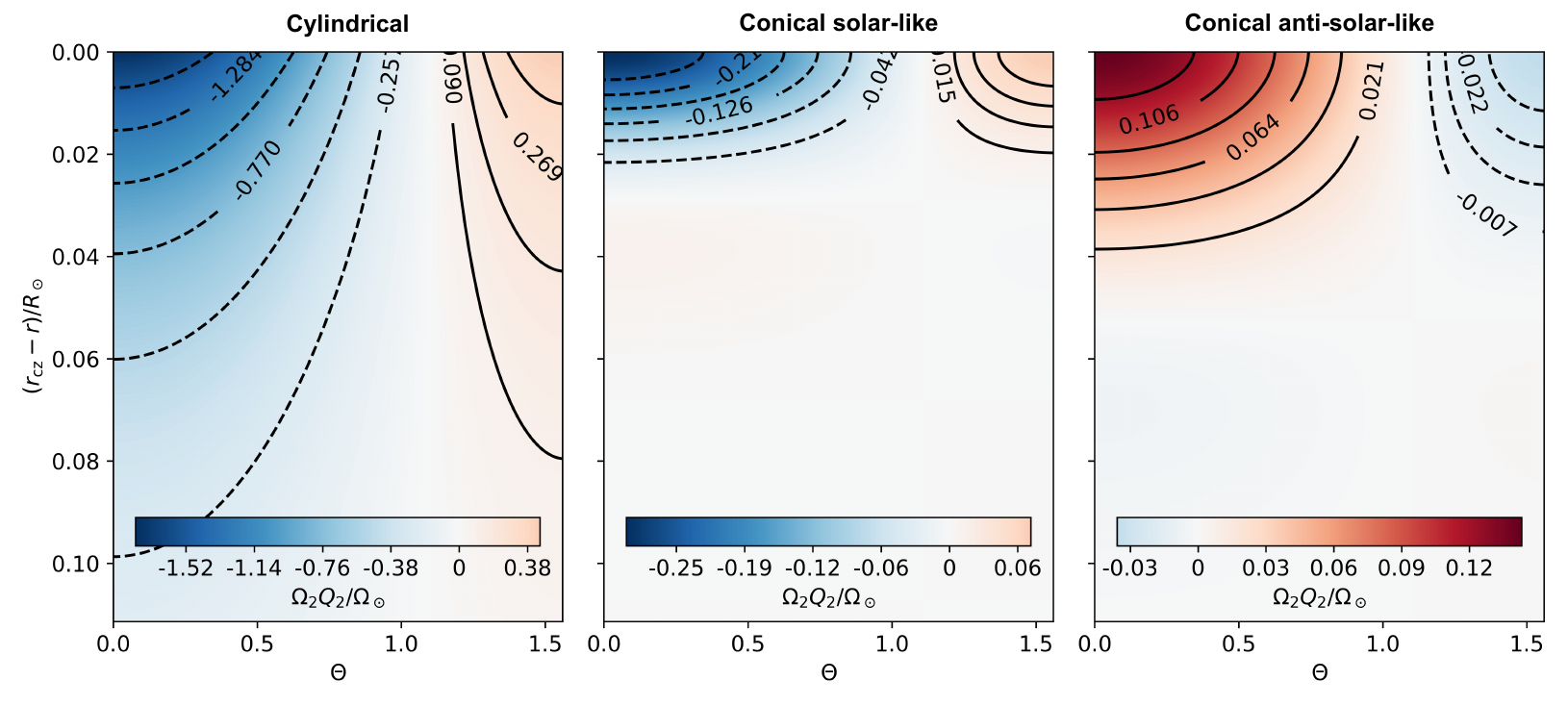} 
  	\caption{Differential rotation in the tachocline $\Omega_2Q_2$ as a function of the colatitude $\theta$ and of the distance to the base of the convective layer (located at $(r_{cz}-r)/R_{\odot}=0)$, computed using the prescription of \cite{Mathisetal2004} for the turbulent horizontal eddy-viscosity. {\bf Left:} $t=1.0\times10^{8}$ years associated with a cylindrical forcing from the convective zone. {\bf Middle:} $t=1.0\times10^{9}$ years and a conical solar-like forcing. {\bf Right:} $t=6.7\times10^{9}$ years and a conical anti-solar-like forcing.}
 	 \label{fig_rot_diff}
\end{figure}

Figure \ref{fig_rot_diff} shows the differential rotation inside the tachocline $\Omega_{2}Q_{2}$ normalised by the solar rotation $\Omega_{\odot}$, as a function of the colatitude $\theta$, of the distance from the base of the convective layer (located at $(r_{cz}-r)/R_{\odot}=0$, with $r$ the radial coordinate, $r_{cz}$ the radius of the base of the convective region and $R_{\odot}$ the solar radius) and at three stages of the evolution, each corresponding to a different forcing imposed by the convective envelope. To express the complete rotation in the tachocline, the differential rotation $\Omega_{2}Q_{2}$ needs to be added to the mean shellular rotation of the layer $\overline{\Omega}$.
During the early main sequence, when the star has a high rotation rate, the differential rotation in the tachocline is important with a fast equator and a slow pole. The latitudinal shears extend far into the radiative interior, reaching up to 10\% R$_{\odot}$. Later on the MS, the star undergoes magnetic braking and the differential rotation is less intense, even in the tachocline. When the forcing from the convection zone is conical solar-like, the tachocline is at its most confined, at about 2\% R$_{\odot}$. At the end of the MS, the differential rotation in the tachocline reverses, with a fast pole and a slow equator, imposed by the overlying conical anti-solar-like rotation of the convective region. While more extended into the interior than in the previous evolution phase, the intensity of the differential rotation is still decreasing, reaching a minimum. 

This model highlighted two key parameters that govern the dynamics observed in figure \ref{fig_rot_diff}:
\begin{itemize} 
	\item $\alpha$, which depends on the mean shellular rotation in the tachocline $\overline{\Omega}$, the (thermal) Brunt-Väisälä frequency in the tachocline $N$, the vertical thermal diffusion $\kappa_{v}$ and the turbulent horizontal eddy-viscosity $\nu_{h}$ as prescribed by \cite{Mathisetal2004}:
	\begin{equation}
		\alpha=\left(\frac{\overline{\Omega}}{N}\right)^{1/2}\left(\frac{\kappa_{v}}{\nu_{h}}\right)^{1/4}
	\end{equation}
	\item $\Delta\Omega$, the latitudinal shear applied at the top of the tachocline by the convective envelope, using the results obtained by \cite{Brunetal2022}.
\end{itemize}
Indeed, the extension of the differential rotation in the interior is entirely determined by $\alpha$ whereas the maximum value imposed at the top of the tachocline is set by a combination of both $\alpha$ and $\Delta\Omega$. 

\section{Transport of chemical elements in the tachocline}
The large-scale meridional circulation and turbulence are prime actors of the transport and mixing of chemical elements in stellar radiative regions and thus also in tachoclines. \cite{chaboyerZahn1992} have shown that the advection-diffusion of the chemical elements can be represented as a simple diffusion due to the erosion of the transport by the anisotropic turbulence. As such, the transport of the chemical elements can be parametrized using an effective turbulent diffusion coefficient D$_\textrm{eff}$.

\begin{figure}[ht!]
\centering
\begin{minipage}[b]{0.36\linewidth}
 \centering
 \includegraphics[width=\textwidth,trim=2cm 8.5cm 2cm 8.5cm,clip]{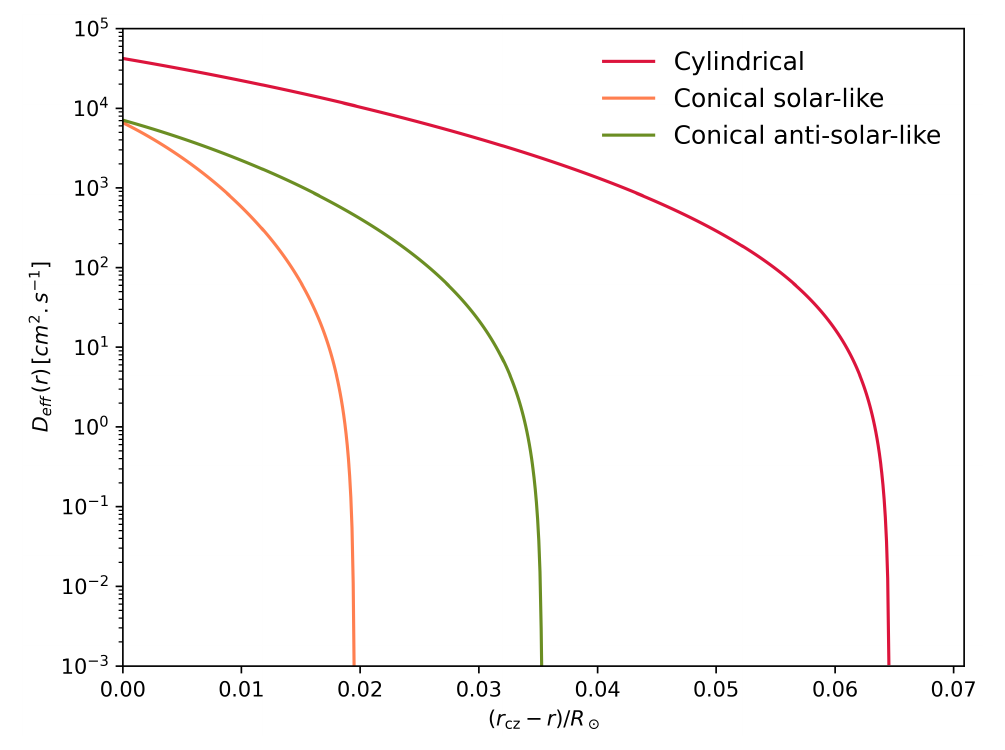}
\end{minipage}
\begin{minipage}[b]{0.35\linewidth}
\centering
  \caption{Effective turbulent diffusion coefficient $D_{\textrm{eff}}$ as a function of the distance to the base of the convection zone $((r_{cz}-r)/R_{\odot}=0)$ for the three times considered in figure \ref{fig_rot_diff}.}
  \label{fig_deff}
  \vspace{0.9 cm}
\end{minipage}
\end{figure}

Figure \ref{fig_deff} shows the profiles of $\textrm{D}_{\textrm{eff}}$ in the tachocline as a function of the distance to the base of the convective envelope, for the three evolutions considered in figure \ref{fig_rot_diff}. During the early MS, $\textrm{D}_{\textrm{eff}}$ is of order $\sim 10^{4}$cm$^{2}$.s$^{-1}$  at the top of the tachocline and decreases slowly into the interior. Note that it still reaches $\sim 10^3$cm$^2$.s$^{-1}$ at 0.04R$_\odot$. Later on the MS, when the forcing is conical solar-like, the chemical mixing is less intense at the border of the convective zone and less extended into the interior, barely reaching 2\% R$_{\odot}$. At the end of the MS, its intensity at the top of the tachocline has the same order of magnitude but gained a bit in depth due to the more extended differential rotation and meridional circulation in this late stage.

For each case, $\textrm{D}_{\textrm{eff}}$ is maximum at the top of the tachocline. The transport of the chemical elements is also more efficient in the early phases of evolution, a result which concurs with \cite{Dumontetal2021}. Indeed, they implemented a turbulent mixing at the tachocline in STAREVOL, following \cite{Brunetal1999}, and showed that for cases where the tachocline is more confined (i.e. for the conical solar-like and anti-solar-like forcings), the mixing is dominated by the overshoot by multiple orders of magnitudes.

\section{Conclusions}

The formalism of \cite{MathisZahn2004} can self-consistently treat hydrodynamical stellar tachoclines all along the evolution of low mass stars, a model to which we have included the evolution of the differential rotation in the convective region as predicted by 3D MHD simulations. The rotation and meridional circulation are significantly modified during the main sequence, the early fast rotating phases showing a thicker tachocline and a more important mixing of the chemical elements by the horizontal turbulence and meridional circulation, both in depth and in intensity. Later on the main sequence, the tachocline is more confined and this mixing is less efficient. Simulation works have shown that, due to this confined state, the overshoot is likely to dominate the tachocline's effective mixing at the base of the convective region \cite[see][]{Dumontetal2021}. However, this highlights the key role of the tachocline in the early main sequence, when the turbulence and meridional circulation are the only processes capable of mixing this deep into the radiative interior. It also shows a different dependence on rotation that goes contrary to the overshooting scheme (i.e. less overshoot for fast rotation, as predicted by \cite{AugustsonMathis2019}).

\begin{acknowledgements}
C.M. and S.M. acknowledge support from the European Research Council (ERC) under the Horizon Europe program (Synergy Grant agreement 101071505: 4D-STAR), from the CNES SOHO-GOLF and PLATO grants at CEA-DAp, and from PNST and PNPS (CNRS/INSU). While partially funded by the European Union, views and opinions expressed are however those of the authors only and do not necessarily reflect those of the European Union or the European Research Council. Neither the European Union nor the granting authority can be held responsible for them. L.A. acknowledges support from the Swiss National Science Foundation (SNF; Project 200021L-231331) and the French Agence Nationale de la Recherche (ANR-24-CE93-0009-01) “PRIMA - PRobing the origIns of the Milky WAy’s oldest stars”.
\end{acknowledgements}

\bibliographystyle{aa}  
\bibliography{Moisset_S20.bib} 

\end{document}